\documentclass[aps,prl,preprint,groupedaddress]{revtex4-1}

\usepackage{graphics}

\begin{document}


\title{Octupole strength in the neutron-rich calcium isotopes}


\author{L. A. Riley}
\affiliation{Department of Physics and Astronomy, Ursinus College, Collegeville, PA 19426, USA}

\author{D. M. McPherson} 
\affiliation{Department of Physics, Florida State University, Tallahassee, FL 32306, USA}

\author{M. L. Agiorgousis} 
\affiliation{Department of Physics and Astronomy, Ursinus College, Collegeville, PA 19426, USA}

\author{T.R. Baugher} 
\affiliation{National Superconducting Cyclotron Laboratory, Michigan State University, East Lansing, MI, 48824, USA}
\affiliation{Department of Physics and Astronomy, Michigan State University, East Lansing, MI, 48824, USA} 

\author{D. Bazin} 
\affiliation{National Superconducting Cyclotron Laboratory, Michigan State University, East Lansing, MI, 48824, USA}

\author{M. Bowry} 
\affiliation{National Superconducting Cyclotron Laboratory, Michigan State University, East Lansing, MI, 48824, USA}
\affiliation{Department of Physics and Astronomy, Michigan State University, East Lansing, MI, 48824, USA}

\author{P. D. Cottle} 
\affiliation{Department of Physics, Florida State University, Tallahassee, FL 32306, USA}

\author{F. G. DeVone} 
\affiliation{Department of Physics and Astronomy, Ursinus College, Collegeville, PA 19426, USA}

\author{A. Gade} 
\affiliation{National Superconducting Cyclotron Laboratory, Michigan State University, East Lansing, MI, 48824, USA}
\affiliation{Department of Physics and Astronomy, Michigan State University, East Lansing, MI, 48824, USA} 

\author{M. T. Glowacki} 
\affiliation{Department of Physics and Astronomy, Ursinus College, Collegeville, PA 19426, USA}

\author{S. D. Gregory} 
\affiliation{Department of Physics and Astronomy, Ursinus College, Collegeville, PA 19426, USA}

\author{E. B. Haldeman} 
\affiliation{Department of Physics and Astronomy, Ursinus College, Collegeville, PA 19426, USA}

\author{K. W. Kemper} 
\affiliation{Department of Physics, Florida State University, Tallahassee, FL 32306, USA}

\author{E. Lunderberg} 
\affiliation{National Superconducting Cyclotron Laboratory, Michigan State University, East Lansing, MI, 48824, USA}
\affiliation{Department of Physics and Astronomy, Michigan State University, East Lansing, MI, 48824, USA} 

\author{S. Noji} 
\affiliation{National Superconducting Cyclotron Laboratory, Michigan State University, East Lansing, MI, 48824, USA}
\affiliation{Joint Institute for Nuclear Astrophysics - Center for the Evolution of the Elements, Michigan State University, East Lansing, MI 48824, USA}

\author{F. Recchia} 
\altaffiliation{Dipartimento di Fisica e Astronomia “Galileo Galilei”, Universit‘a degli Studi di Padova, I-35131 Padova, Italy} 
\affiliation{National Superconducting Cyclotron Laboratory, Michigan State University, East Lansing, MI, 48824, USA}

\author{B. V. Sadler} 
\affiliation{Department of Physics and Astronomy, Ursinus College, Collegeville, PA 19426, USA}

\author{M. Scott} 
\affiliation{National Superconducting Cyclotron Laboratory, Michigan State University, East Lansing, MI, 48824, USA}
\affiliation{Department of Physics and Astronomy, Michigan State University, East Lansing, MI, 48824, USA} 

\author{D. Weisshaar} 
\affiliation{National Superconducting Cyclotron Laboratory, Michigan State University, East Lansing, MI, 48824, USA}

\author{R. G. T. Zegers} 
\affiliation{National Superconducting Cyclotron Laboratory, Michigan State University, East Lansing, MI, 48824, USA} 
\affiliation{Department of Physics and Astronomy, Michigan State University, East Lansing, MI, 48824, USA} 
\affiliation{Joint Institute for Nuclear Astrophysics - Center for the Evolution of the Elements, Michigan State University, East Lansing, MI 48824, USA}


\date{\today}

\begin{abstract}
Low-lying excited states of the neutron-rich calcium isotopes
$^{48-52}$Ca have been studied via $\gamma$-ray spectroscopy following
inverse-kinematics proton scattering on a liquid hydrogen target using
the GRETINA $\gamma$-ray tracking array. The energies and strengths of
the octupole states in these isotopes are remarkably constant,
indicating that these states are dominated by proton excitations.
\end{abstract}


\maketitle

\section{I. Introduction}

Insights about nuclear structure can often be found by examining the
systematic behavior of a series of isotopes or isotones, and the
advent of intense beams of exotic isotopes has opened new
opportunities for such systematic studies. In the present work, we
examine the systematic behavior of both octupole and quadrupole modes
in the neutron-rich isotopes of calcium by means of inelastic
scattering of protons in inverse kinematics. We report the results 
of inelastic proton scattering experiments on $^{49,51,52}$Ca, and
combine these results with previous results on
$^{48,50}$Ca~\cite{Ril14} to examine quadrupole and octupole
excitations in the A=48-52 Ca isotopes. While the energies of the
$2^+_1$ states in $^{48,50,52}$Ca, which have large neutron excitation
components, vary dramatically with neutron number, the energies of the
$3^−_1$ states in these isotopes are remarkably constant, demonstrating
that these octupole states are dominated by proton excitations. 

Furthermore, the inelastic proton scattering reaction on the odd-N
isotopes $^{49,51}$Ca allows us to identify members of multiplets that
result from coupling of the odd neutron (in the case of $^{49}$Ca) or
neutron hole (in the case of $^{51}$Ca) to the octupole phonons in
$^{48,52}$Ca. The deformation lengths for scattering to the multiplet
members in $^{49,51}$Ca can be explained in a simple weak-coupling
picture, once again confirming the dominant proton character of the
octupole phonons in the core nuclei.

\section{II. Experiment}

The experiment was performed at the Coupled-Cyclotron Facility of the
National Superconducting Cyclotron Laboratory at Michigan State
University. A cocktail beam was produced by the fragmentation of a
130~MeV/u $^{76}$Ge primary beam in a 376~mg/cm$^2$ $^9$Be production
target. The secondary products were separated by the A1900 fragment
separator~\cite{A1900}. The momentum acceptance of the A1900 was
3\%. A 45~mg/cm$^2$ aluminum achromatic wedge was used to enhance
separation of the cocktail by $Z$.

Secondary beam particles were identified by energy loss in a silicon
surface barrier detector and by 
their time of flight.  The beam then traversed the Ursinus College
Liquid Hydrogen Target, based on the design of Ryuto et
al.~\cite{Ryu05}. The target was installed at the pivot point of the
S800 magnetic spectrograph~\cite{S800}. Projectile-like reaction
products were identified by energy loss in the S800 ion chamber and
time of flight.  The secondary beam, composed of products spanning the
range $14 \leq Z \leq 23$, included $^{49,51,52}$Ca, the subjects of
the present work. Total numbers of beam particles, average rates, and
mid-target energies are given in Table~\ref{tab:beam}. 

\begin{table}
\caption{\label{tab:beam} Beam yields, average rates, and mid-target
  kinetic energies} 
\begin{ruledtabular}
\begin{tabular}{cccc}
         & Total Beam Particles & Rate  & Mid-target KE \\
         &                      & [pps] & [MeV/u] \\
\hline\hline
$^{49}$Ca & $3.7 \times 10^7$ & 90   & 94 \\
$^{51}$Ca & $5.1 \times 10^6$ & 12   & 86 \\
$^{52}$Ca & $3.3 \times 10^5$ & 0.81 & 84 \\
\end{tabular}
\end{ruledtabular}
\end{table}

The liquid hydrogen target consisted of a cylindrical aluminum target
cell with 125~$\mu$m Kapton entrance and exit windows mounted on a
cryocooler. The nominal target thickness was 30~mm.  The target cell
and cryocooler were surrounded by a 1~mm thick aluminum radiation
shield with entrance and exit windows covered by 5~$\mu$m aluminized
Mylar foil. A resistive heater mounted between the cryocooler and the
target cell was used to maintain the temperature and pressure of the
target cell at 16.00(25)~K and 868(10)~Torr throughout the
experiment. The variation in the temperature and pressure of the
target cell correspond to a 0.3~\% uncertainty in target density.

\begin{figure}
\scalebox{0.48}{
  \includegraphics{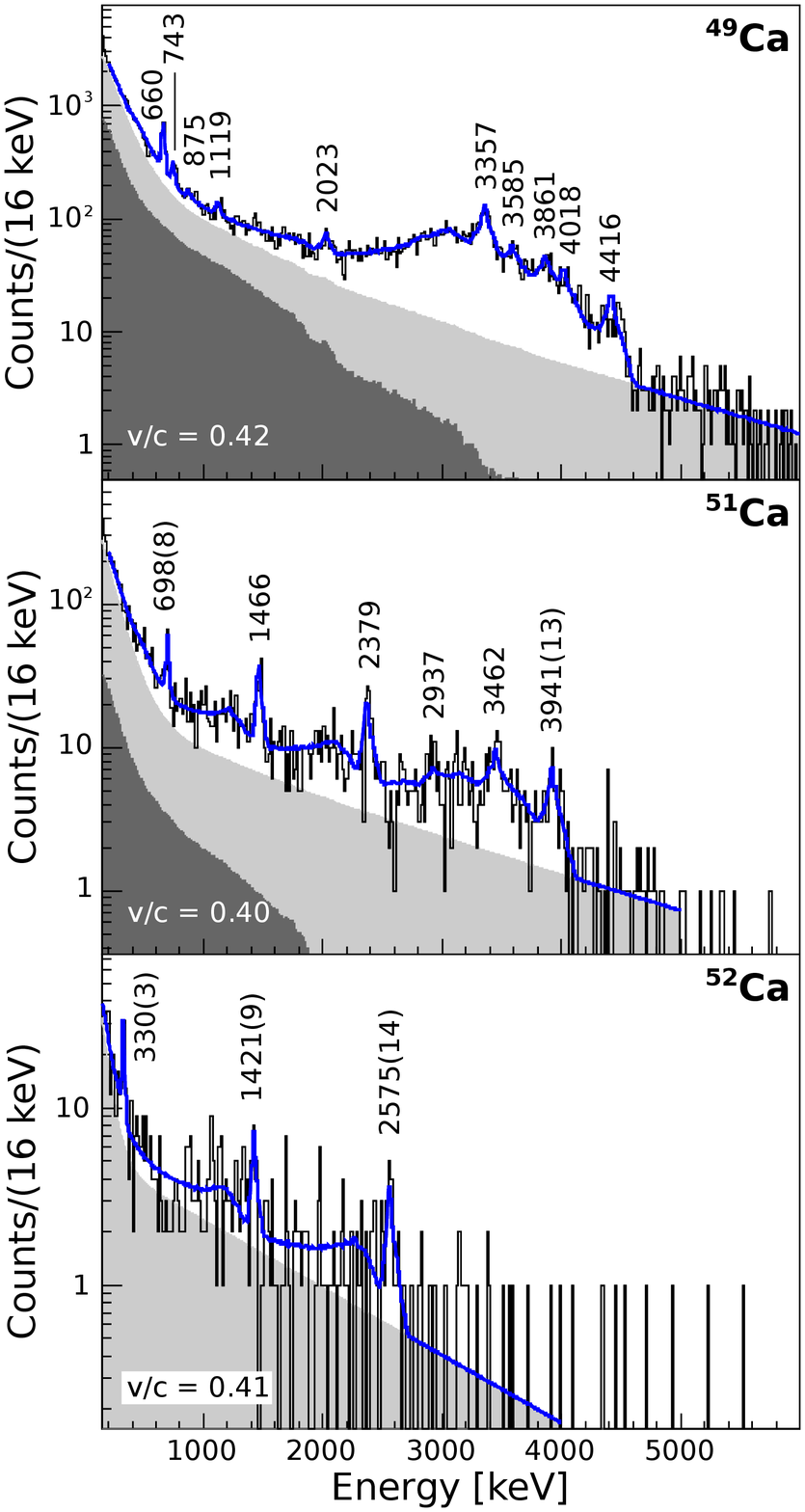}
}
\caption{\label{fig:spectra} Projectile-frame $\gamma$-ray spectra
  measured in coincidence with incoming and outgoing $^{49,51,52}$Ca
  particle-identification gates. The solid curves are the GEANT4
  fits. The shaded region is the background, consisting of nonprompt
  (dark gray) and prompt (light gray) components.}
\end{figure}

\begin{table*}
\caption{\label{tab:gammas} Level energies, spins and parities,
  and $\gamma$-ray energies from Refs.~\cite{Bur08, Bro06,
    For08, Gad06}, $\gamma$-ray energies, intensities relative
  to that of the $2^+_1 \rightarrow 0^+_\mathrm{g.s.}$ transitions,
  branching ratios (BR), and cross sections from the present
  work.}
\begin{ruledtabular}
\begin{tabular}{crcrccrccc}
&$E_\mathrm{level}$ [keV] & $J^\pi$ [$\hbar$] & $E_\gamma$ [keV] &BR [\%]&
  $J^\pi$ [$\hbar$] & $E_\gamma$ [keV] & $I_\gamma$ [\%] & BR [\%] & $\sigma$ [mb]
\\\hline\hline
$^{49}$Ca  &\multicolumn{4}{c}{Refs.~\cite{Bur08, Bro06}} && & & \\\cline{3-5}
&2023.2(3) & $1/2^-$     & 2023.12(26)&     &&  2023 &  12(3) & & 0.6(1) \\
&3354.7(6) & $7/2^-$     & 3356.7(10) &     &&  3357 & 100(7) & & 1.6(3) \\
&3585.0(8) & $5/2^-$     & 3585.0(8)  &     &&  3585 &  29(6) & & 1.4(3) \\
&3861(2)   &$(1/2^-,3/2^-)$& 3859.7(9)&     &&  3861 &  34(6) & & 1.4(3) \\
&4013.6(6) & $9/2^+$     & 4017.5  & 9.2(8) &&  4018 &  29(14)& & 3.3(8) \\
&          &             &  660.3  &  84(1) &&   660 &  67(6) & &        \\
&          &             &  150.9  &  6.7(8)&&   --- &        & &        \\
&4416(2)   & $5/2^+$     &  4416   &        &&  4416 &  30(3) & & 1.5(7) \\
&4757.0(10)& $(5/2^+)$   &  743.3  &        &$(3/2^+)$&   743 &  20(3) & & 1.0(1) \\
&4885(3)   & $9/2^+$     & 1531    &  20(6) &&   --- &        & & 0.2(1) \\
&          &             &  875    &  80(23)&&   875 &   4(2) & &        \\
&5132.8(10)&             & 1119    &        && 1119 &   9(2) & & 0.5(1) \\\hline
$^{51}$Ca  &\multicolumn{4}{c}{Ref.~\cite{For08}} && & & \\\cline{3-5}
&2379     & $(5/2^-)$    & 2379 &    && 2379  &100(13)&       & $<0.9$  \\
&2937     & $(3/2^-)$    & 2937 &    && 2937  & $<22$ &       & $<0.8$  \\
&3462     & $(7/2-)$     & 3462 &    && 3462  & 66(18)&       & $<1.0$  \\
&3845     & $(7/2^+)$    & 1466 &  &$(7/2^+)$& 1466  & 93(18)&       & 3.4(5) \\
&3941(10) &              &      &  &$(5/2^+)$&3941(13)&82(16)&       & 2.9(4) \\
&4155     & $(9/2^+)$    & 693  &  &$(9/2^+)$& 698(8)& 52(11)&       & 1.9(3) \\\hline
$^{52}$Ca  &\multicolumn{4}{c}{Ref.~\cite{Gad06}} && & & \\\cline{3-5}
&2563.1(10) & $2^+$ & 2563.1(10) & &       & 2575(14) & 100(23)& & $<6$ \\
&3990 & $(3^-)$ & 1427(1)        & &$(3^-)$& 1421(9)  & 80(31)  & & 9(3) \\
&     &         &                & &       &  330(3)  & 67(23)  & &  \\
\end{tabular}
\end{ruledtabular}
\end{table*}

\begin{figure*}
\scalebox{0.3}{
  \includegraphics{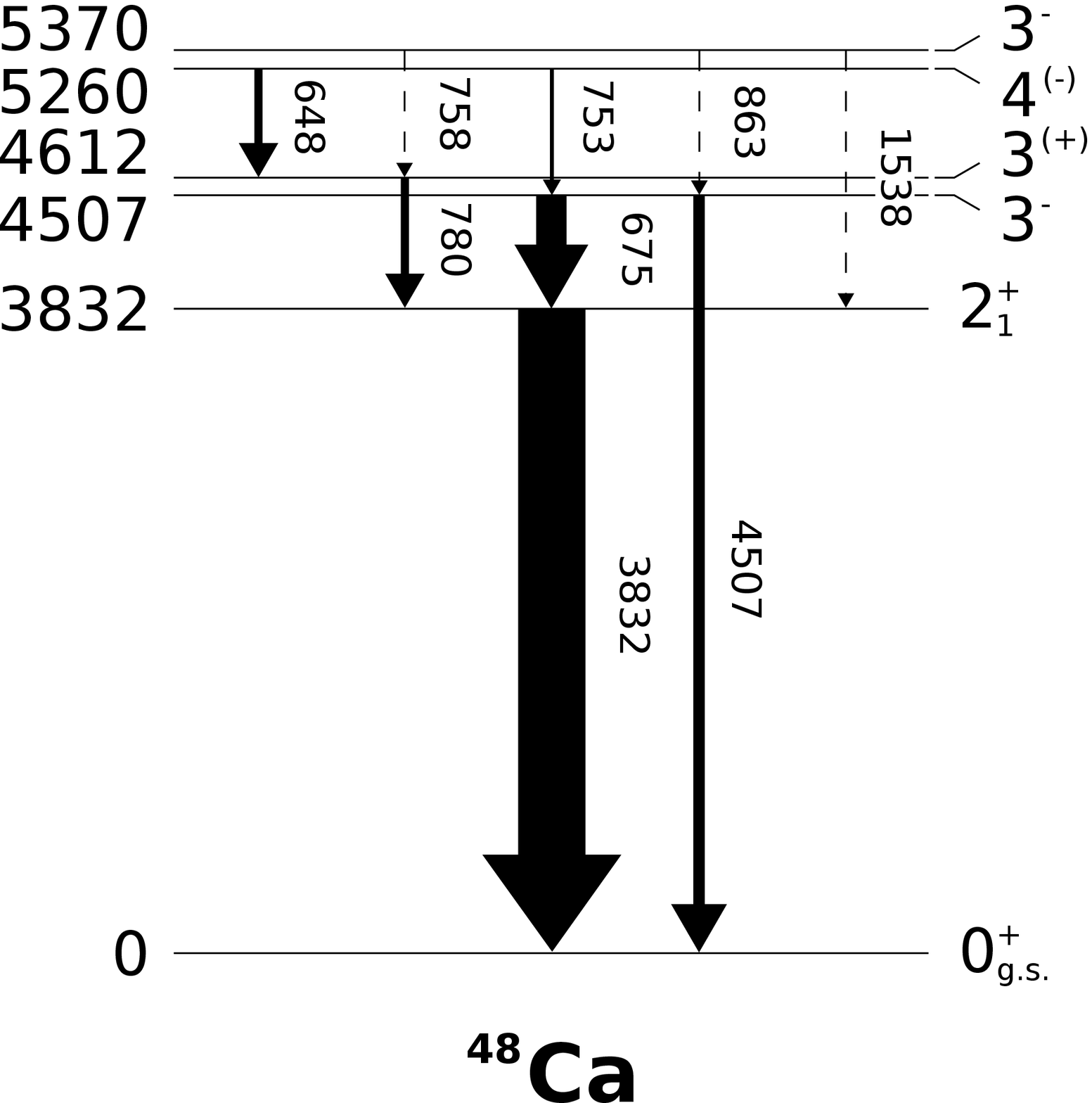}
  \hspace{2cm}
  \includegraphics{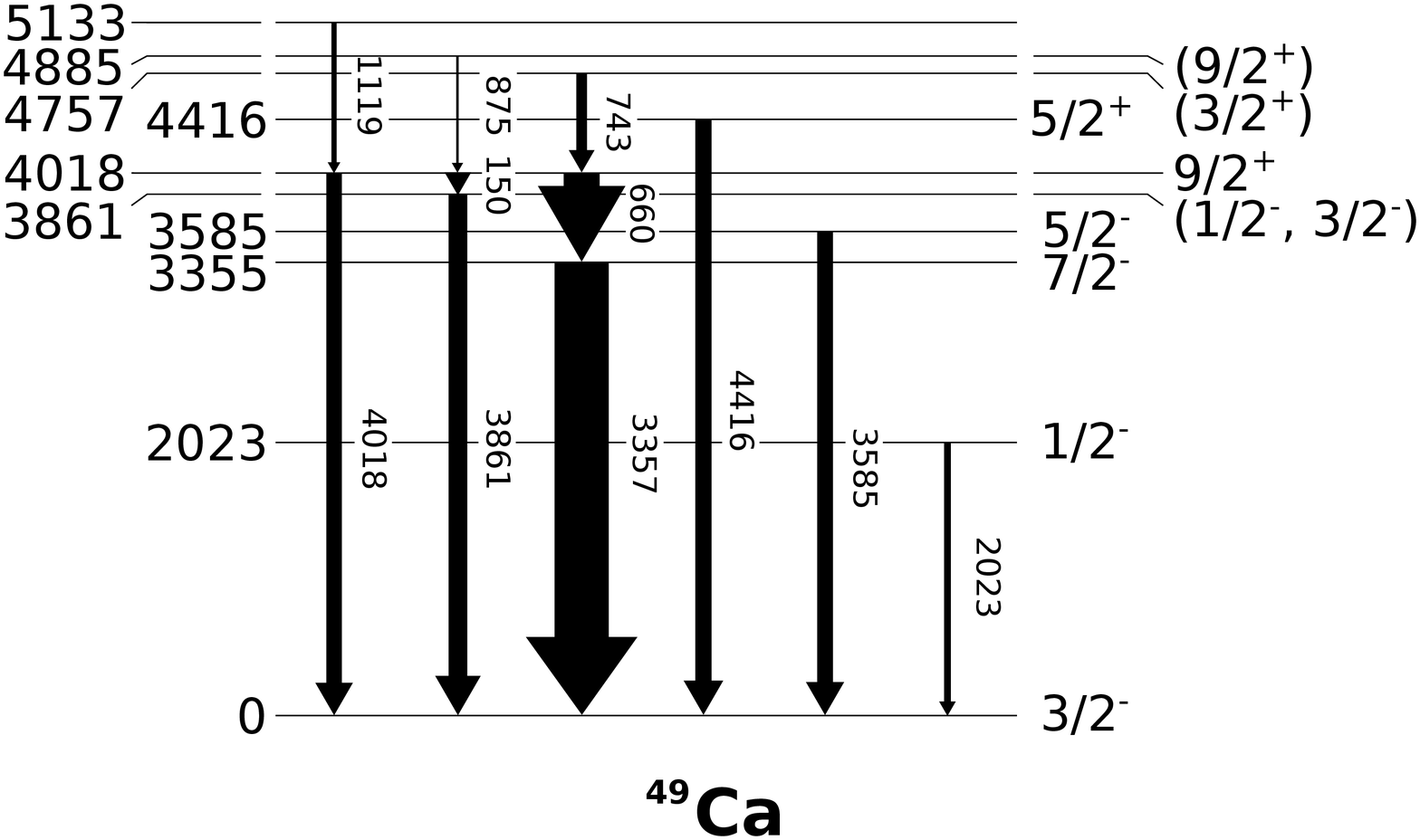}
}\\
\vspace{1 cm}
\scalebox{0.3}{
  \includegraphics{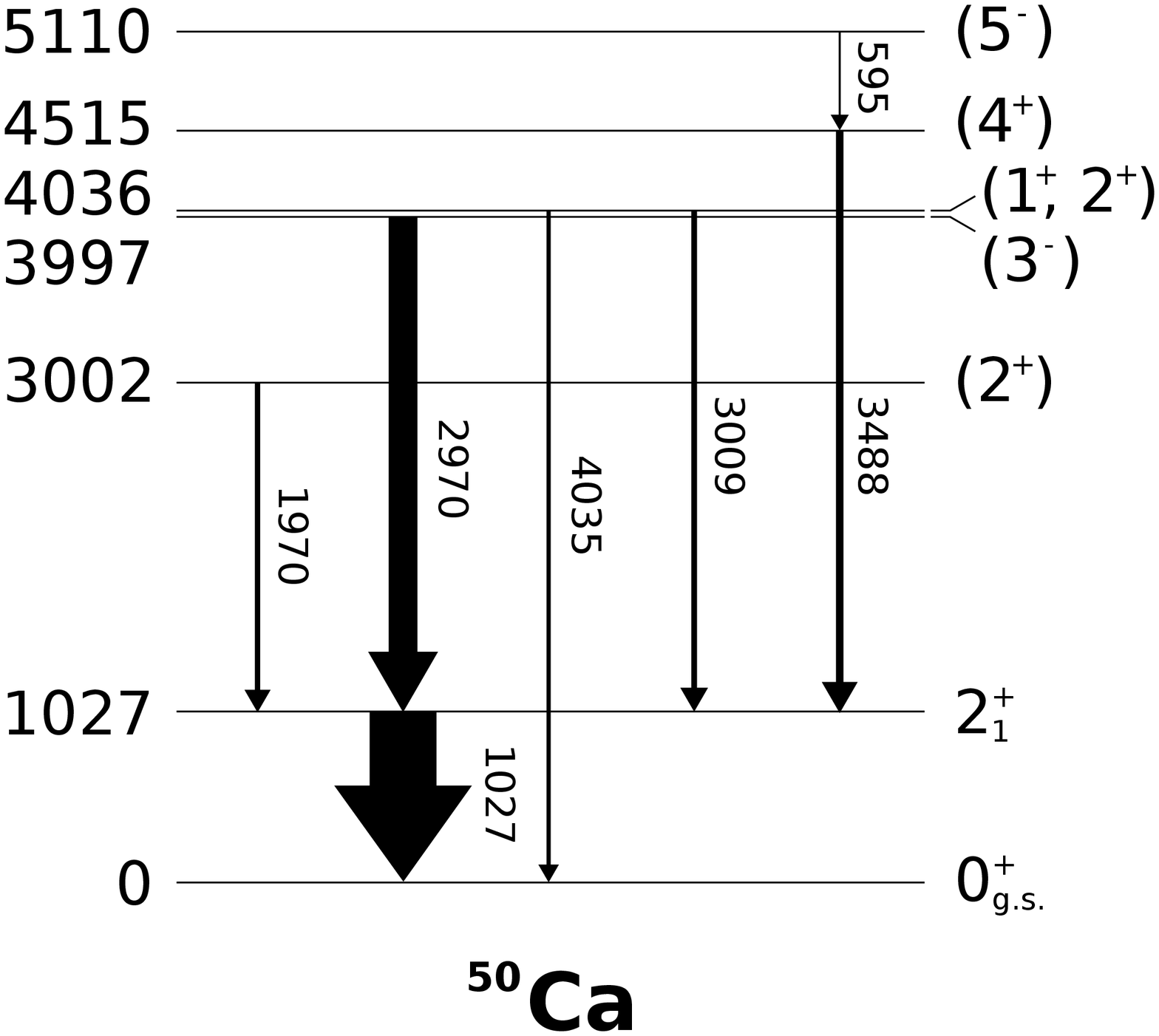}
  \hspace{2cm}
  \includegraphics{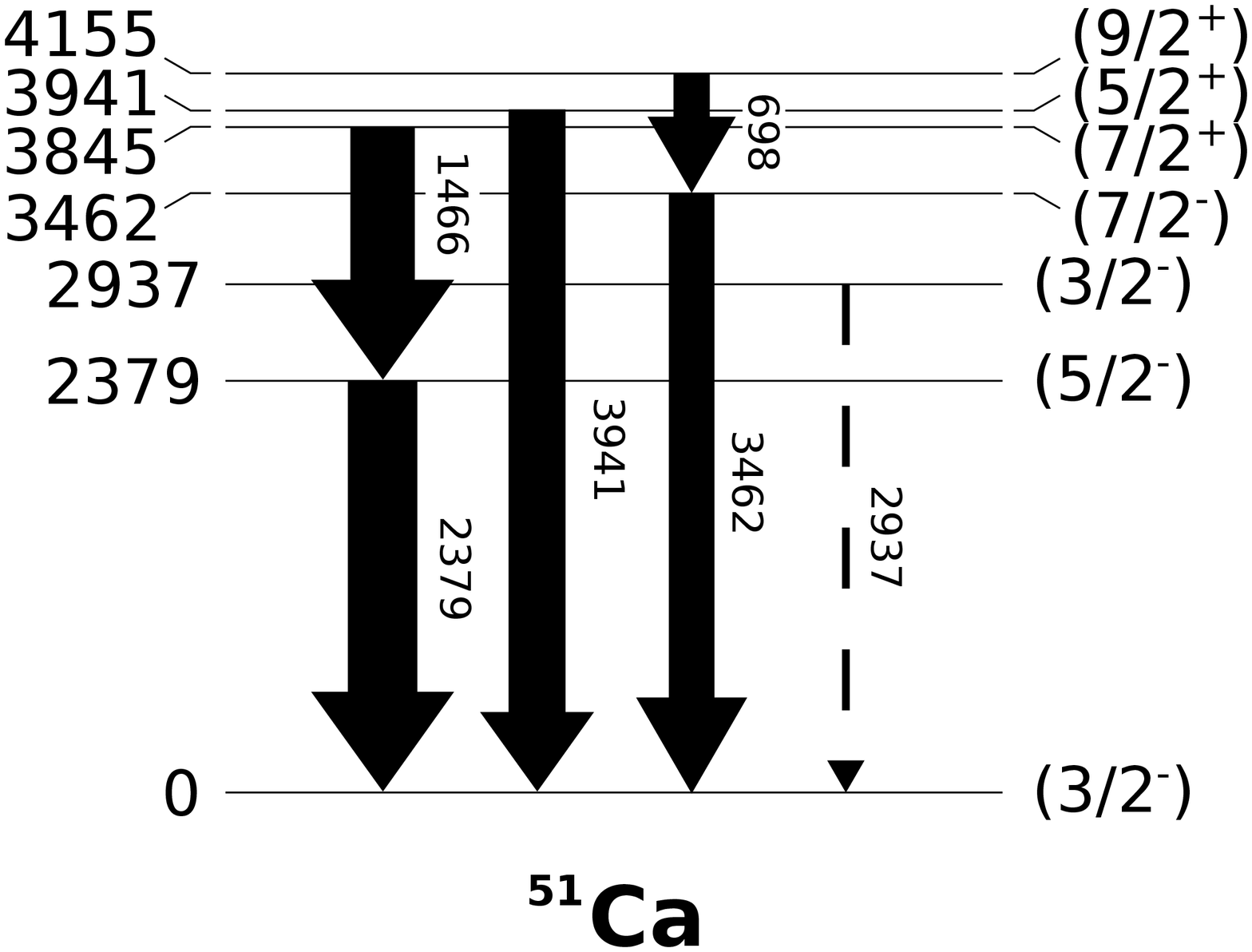}
  \hspace{2cm}
  \includegraphics{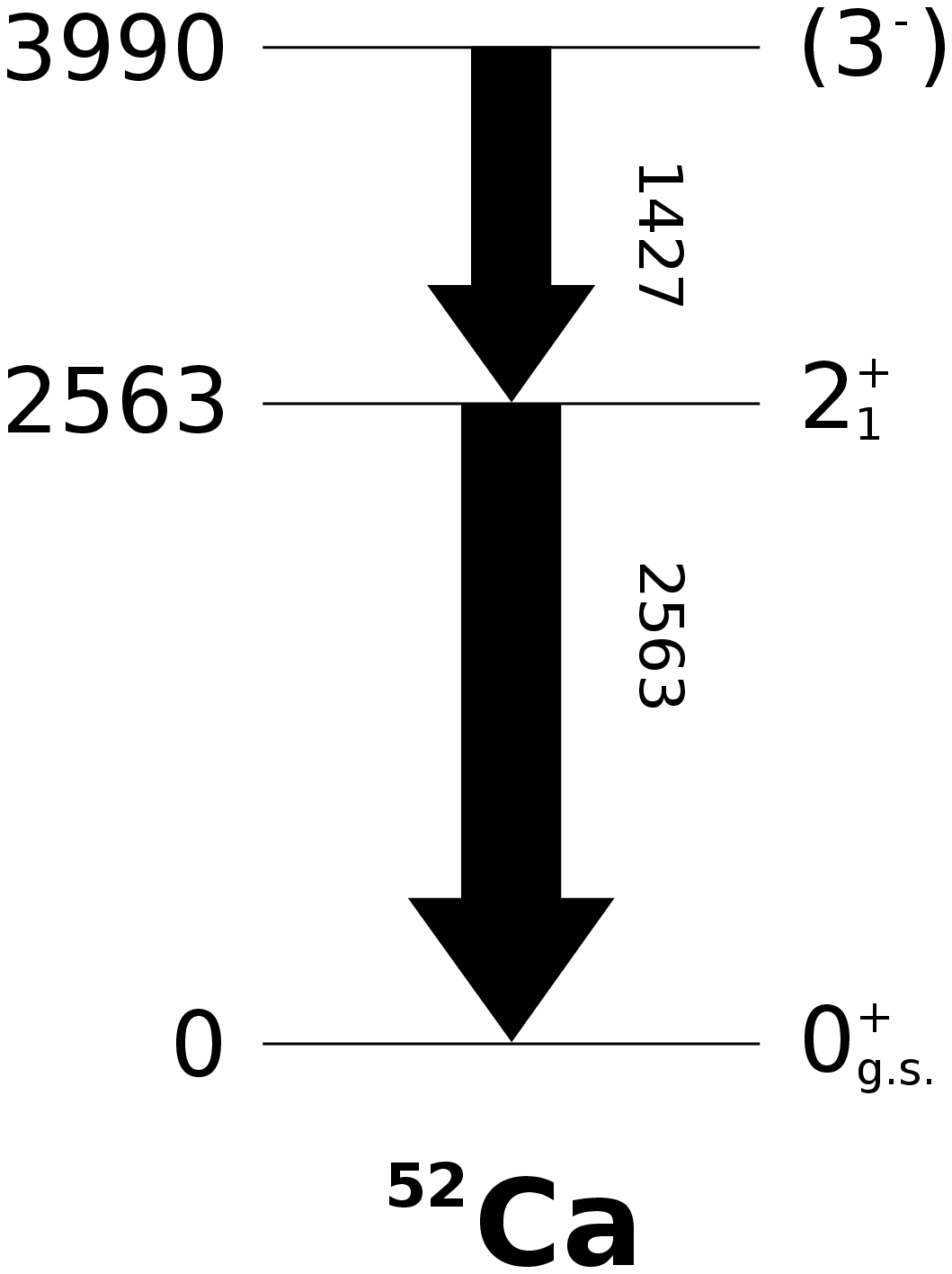}
}
\caption{\label{fig:levels} Partial level schemes of $^{48-52}$Ca
  showing levels populated in the present work. Arrow widths are
  proportional to the measured $\gamma$-ray intensities.}
\end{figure*}

The GRETINA $\gamma$-ray tracking array~\cite{GRETINA} was installed
in the configuration compatible with the liquid hydrogen target
described in Ref.~\cite{Ril14}.  Projectile-frame $\gamma$-ray spectra 
of $^{49,51,52}$Ca measured in coincidence with incoming and outgoing
particle-identification gates corresponding to inverse-kinematics
proton scattering are shown in Fig.~\ref{fig:spectra}.  

\section{III. Analysis and Results}

Intensities of
measured $\gamma$ rays were extracted by fitting
\textsc{geant4}~\cite{GEANT4} simulations to the measured
projectile-frame $\gamma$-ray spectra as described in
Ref.~\cite{Ril14}. The fits are the smooth curves in
Fig.~\ref{fig:spectra}.  Level energies, spins and parities, and
$\gamma$-ray energies from Refs.~\cite{Bur08, Bro06, For08, Gad06},
$\gamma$-ray energies and relative intensities, branching ratios (BR),
and cross sections for $^{49,51,52}$Ca from the present work are
listed in Table~\ref{tab:gammas}.  The states of $^{48-52}$Ca
populated in the present experiment are illustrated in
Fig.~\ref{fig:levels}. The $\gamma$-ray intensities, branching ratios,
cross sections, and levels in $^{48,50}$Ca observed in this reaction
were reported previously~\cite{Ril14}.

Next, we examine the $J^\pi$ assignments for the three radioactive nuclei for 
which $(p,p')$ data are reported here for the first time --- $^{49,51,52}$Ca.

In $^{49}$Ca, the ground state ($J^\pi=3/2^-$) and the first excited state at
2023 keV ($J^\pi=1/2^-$) have been established as the $p_{3/2}$ and
$p_{1/2}$ single neutron states via the $(d,p)$ reaction.  The
significant $(p,p')$ cross sections seen here for the 3355 keV
($J^\pi=7/2^-$) and 3585 keV ($J^\pi=5/2^-$) states support a picture
in which these states arise from the coupling of the $p_{3/2}$ neutron
to the $2^+_1$ state in the $^{48}$Ca core nucleus.  The 
$J^\pi=7/2^-$ assignment for the 3355~keV state was made by Montanari et
al.~\cite{Mon11, Mon12} on the basis of $\gamma$-ray angular distribution and
polarization measurements of the ground state transition with a
heavy-ion transfer reaction.  The 3585~keV state is populated weakly
($C^2S=0.11$) in the $(d,p)$ reaction~\cite{Bur08}, but the proton angular
distribution in that reaction clearly indicates a $L=3$ transfer.
Therefore, the state's wavefunction includes a fragment of the
$f_{5/2}$ single neutron state, and the state has $J^\pi=5/2^-$.

The 3861 keV state also has a significant $(p,p')$ cross section, so it
is likely to be a member of either the $2^+_1$ core state multiplet or the
multiplet arising from the coupling of the $p_{3/2}$ odd neutron to the
core octupole vibration state.  The data on this state available from
$(d,p)$ are ambiguous, and there is no information available on
$\gamma$-decays from this state.  However, Navon et al.~\cite{Nav79}
determined that the isobaric analog resonance in $^{49}$Sc has $L=1$,
so that the wavefunction of this state must have an admixture of a
$p_{3/2}$ or $p_{1/2}$ single neutron and the state has either
$J^\pi=3/2^-$ or $J^\pi=1/2^-$.

The next state in $^{49}$Ca seen in the present experiment is at
4014~keV and has the largest cross section in this isotope.  This
state was observed in $(d,p)$ with $L=4$, indicating that it has an
admixture of the $g_{9/2}$ single neutron state and, therefore,
$J^\pi=9/2^+$.  Montanari et al.~\cite{Mon11, Mon12} confirmed this using
angular distribution and polarization data for the 660~keV
$\gamma$-ray.  Therefore, we deduce that this state is the highest
spin member of the multiplet resulting from the coupling of the
$p_{3/2}$ odd neutron to the $3^-_1$ state in the $^{48}$Ca core.

The 4416~keV state, which also has a significant cross section in the
present $(p,p')$ experiment, was observed strongly enough in $(d,p)$ to
determine an $L=2$ transfer~\cite{Bur08} and therefore $J^\pi=3/2^+$
or $5/2^+$, and the analyzing power data from the same experiment give
$J^\pi=5/2^+$ definitively.  Therefore, this state appears to be a
member of the octupole multiplet.  The 4757~keV state has a
significant $(p,p')$ cross section, but the $(d,p)$ study was less
conclusive regarding $L$ transfer to this state, giving a tentative
$L=2$ result.  This state probably has positive parity and $J=3/2$ or
$5/2$.  As in the case of the 4416~keV state, it is attractive to
assign this state to the octupole multiplet because of its cross
section and likely positive parity.  If it is a member of this
multiplet, then it has $J^\pi=3/2^+$.

The 4885~keV state is weakly populated in $(p,p')$.  While this state is
also weakly populated in $(d,p)$, there was enough information to
determine $L=4$ and, therefore, a likely $J^\pi=9/2^+$ assignment.
Finally, the 5133~keV state was populated significantly here and has
been seen in other reactions, but there is not sufficient information
to make even a tentative $J^\pi$ assignment.

It is likely (though not rigorously established) that the ground state
of $^{51}$Ca has $J^\pi=3/2^-$ since $p_{3/2}$ is the lowest valence
neutron orbit in $^{49}$Ca and this orbit is likely filling in the
$^{49-52}$Ca isotope chain. The present $(p,p')$ measurement of
$^{51}$Ca identified three strong states at 3845, 3941 and 4155~keV.
The strongest states observed in $^{48,49}$Ca(p,p') are octupole vibration
states, so it seems quite likely that the three strong states observed
here in $^{51}$Ca are associated with an octupole vibration as well ---
either the coupling of a $p_{3/2}$ neutron hole to the $3^-_1$ state
in $^{52}$Ca or a $p_{3/2}$ neutron to the corresponding state in
$^{50}$Ca.  This multiplet should include states with $J^\pi=3/2^+$,
$5/2^+$, $7/2^+$ and $9/2^+$. Regardless of the specific spin
assignments of these three strong states in $^{51}$Ca, the energy
centroid of these three state, 3.95(1) MeV, is approximately the
same as the energies of the octupole states in $^{48,49}$Ca. 

In a simple weak-coupling model, the smallest cross section among
these four states would be that for the smallest $J$, in this case the
$3/2^+$ state.  So we can tentatively conclude that the three states
observed here are the $5/2^+$, $7/2^+$ and $9/2^+$ members of the
multiplet.  It is certainly possible that each of these multiplet
members decays to the ground state via $E3$ transitions, as the
$9/2^+$ octupole multiplet member in $^{49}$Ca does.  However, it is
generally more likely that these states decay via $E1$, $M1$ or $E2$
transitions.  Among the 3845, 3941 and 4155~keV states in $^{51}$Ca, only
the 3941~keV state decays directly to the ground state.  Therefore, it
is most likely that this state is the $5/2^+$ member of the octupole
multiplet, decaying via an $E1$ transition.

The most likely $J^\pi$ assignments for the 3845 and 4155 keV states
depend on the assignments for the states to which they $\gamma$-decay
at 2379 and 3462~keV.  Both of these latter states are only weakly
populated in the present $(p,p')$ reaction, so it is unlikely that
they are members of the octupole multiplet.  Furthermore, coupling to
the an octupole vibration is the lowest energy way to produce a
positive parity state in $^{51}$Ca, so the 2379 and 3462~keV states
are likely to be of negative parity.  Both of these states decay
directly to the ground state, which we have tentatively assigned to be
$J^\pi=3/2^-$. So they may have $J^\pi=1/2^-$, $3/2^-$, $5/2^-$, or
$7/2^-$.  The authors of a report on the $\beta$-decay of 
$^{51}$K~\cite{Per06} argue for a $7/2^-$ assignment for the 3462~keV 
state on the basis of their results.  

The 4155~keV state decays exclusively to the 3462~keV state.  If the
3462~keV state is $J^\pi=7/2^-$, then the 4155~keV state is likely
$J^\pi=5/2^+$, $7/2^+$ or $9/2^+$ because $E1$ is the most likely
parity-changing transition. But if the 4155~keV had $J<9/2$, it would
likely decay to one of the lower-energy states, which are probably
lower-spin positive-parity states, because of the $E_\gamma^3$ dependence of
the $E1$ transition probability. So we tentatively assign the 4155~keV
state to be $J^\pi=9/2^+$.  The 3845~keV member of the octupole
multiplet decays to the 2379~keV state, so we tentatively assign
$J^\pi=7/2^+$ for that state. 

In $^{52}$Ca, the cross section for populating the 3990 keV state --- which was 
observed by Gade et al.~\cite{Gad06} via the two-proton knockout reaction 
from $^{54}$Ti and tentatively assigned $J^\pi=3^-$ --- is strong, as is the 
case for the $3^-$ states in $^{48,50}$Ca.  Thus, the present $(p,p')$ 
result provides additional support for the tentative $3^-$ assignment 
for the 3990 keV state in $^{52}$Ca by Gade et al.

It is worth noting that a strong 330 keV $\gamma$ ray was observed in
coincidence with $^{52}$Ca residues in the present study.  It has not
been seen in other reactions, and we are unable to place it in the
level scheme.  We can rule out the possibility that this $\gamma$ ray
feeds the $2^+_1$ state directly, because the intensity of the
2563~keV $2^+_1 \rightarrow 0^+_{g.s}$ transition is significantly
below the combined intensities of the 330~keV and 1427~keV $\gamma$
rays. While we cannot rule out the possibility that the 330~keV
$\gamma$ ray feeds the state at 3990~keV, we consider it to be
unlikely. The intensity of this $\gamma$ ray indicates that the state
it de-excites is highly collective. 
Proton-scattering cross sections measured in nearby neutron-rich
calcium isotopes~\cite{Fuj88,Ril14} suggest that such a state would have
$J^\pi = 3^-$ or $2^+$. However, in either of these cases, we would
expect there to be a strong $\gamma$-ray branch to the $2^+_1$
state, which we do not observe. It is possible that this 330 keV
$\gamma$-ray may feed an isomer in $^{52}$Ca so that we would be
unable to connect it to either of the other states in that nucleus
observed here. Such a scenario has been observed in
$^{48}$Ca, in which a collective $5^-_1$ state 
populated in proton scattering de-excites via a strong $E1$
transition to the long-lived $4^+_1$ state.~\cite{Ril14}

\section{IV. Discussion}

With the present results on $^{49,51,52}$Ca and the previous results 
on $^{48,50}$Ca, we can examine the systematic behavior of quadrupole and
octupole states in these isotopes.  It is remarkable how quickly 
the $2^+_1$ state drops from a high energy
characteristic of a doubly-magic nucleus in $^{48}$Ca (3832~keV) down to
1027~keV in $^{50}$Ca, and then back up to 2563~keV in $^{52}$Ca.
This behavior reflects the isolation of the $p_{3/2}$ neutron orbit
between the $N=28$ major shell closure and the strong $N=32$ subshell
closure, and the dominance of neutron components in the $2^+_1$ states
of these nuclei. 

In contrast, the $3^-_1$ states in $^{48,50,52}$Ca are close in energy
(4507, 3997, and 4318~keV, respectively).  Montanari et
al.~\cite{Mon11, Mon12} argued on the basis of an RPA calculation of the
$3^-_1$ state in $^{48}$Ca that one proton-one hole excitations
dominate the wavefunction of this state. This would imply that adding
neutrons to $^{48}$Ca would have little impact on the energy of the
octupole state and would explain why the $3^-_1$ states in
$^{48,50,52}$Ca are at similar energies.  

Furthermore, the proton dominance of the octupole states in $^{48,50,52}$Ca
would imply that the odd neutron in $^{49}$Ca or $^{51}$Ca would be a
spectator to the octupole excitation of the core, which is the
definition of weak coupling.  The octupole energies shown in Figure 3a for 
the odd-A isotopes $^{49,51}$Ca are the centroids of the octupole states
reported here for these nuclei.  These centroids fit well into the
systematic behavior of the $3_1^-$ states in the even-A isotopes, as they
should in a weak-coupling picture.     

\begin{figure}
\scalebox{0.5}{
  \includegraphics{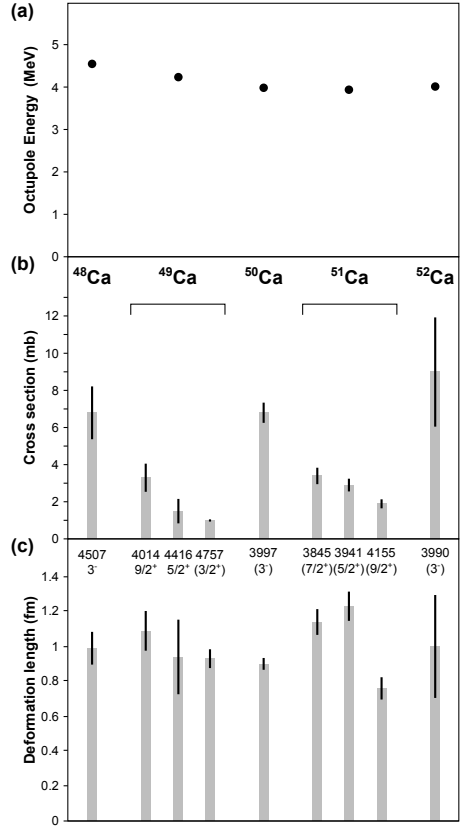}
}
\caption{\label{fig:e3_delta3} (a) Energies (b) proton-scattering
  cross sections, and (c) proton-scattering deformation lengths
  of octupole excitations in $^{48-52}$Ca.}
\end{figure}

Of course, proton dominance of the octupole excitations in the even-$A$
isotopes and weak coupling in the odd-$A$ isotopes should also be
evident in the $(p,p')$ excitation strengths.
Fig.~\ref{fig:e3_delta3}b shows that the $(p,p')$ cross sections to
the $3^-_1$ states in $^{48,50,52}$Ca are identical, to within
experimental uncertainties. The cross sections to the octupole
multiplet members in $^{49,51}$Ca have much smaller cross sections,
but that is to be expected, because in the weak-coupling model the
strength of the octupole vibration of the core nucleus is distributed
among the multiplet members in the odd-$A$ nucleus.  However, the sum
of the cross sections of the individual octupole states 
in $^{49}$Ca, $5.8(11)$ mb, is equal to that of the $3_1^-$ state in its
weak-coupling core nucleus, $^{48}$Ca.  The same is true in $^{51}$Ca:  The
sum of the octupole state cross sections, $8.2(7)$ mb, is equal to
the $3_1^-$ state cross section in its weak-coupling core 
nucleus, $^{52}$Ca.

The deformation lengths extracted for the octupole states in
$^{48-52}$Ca using the coupled-channels code ECIS95~\cite{Ray95} and
the global optical potential of Ref.~\cite{Kon03} are 
shown in Fig.~\ref{fig:e3_delta3}c.  ECIS does not allow for the input
of non-integer spins, so to analyze the cross sections in the odd-A
isotopes, the experimental cross sections were first multiplied by the
factor $(2\lambda+1)(2j+1)/(2I+1)$, which is taken from the weak
coupling result in~\cite{Boh75}.  Here, $\lambda$ is the angular
momentum of the phonon (in this case 3), $j$ is the angular momentum
of the single nucleon being coupled to the phonon (here $j=3/2$), and
$I$ is the angular momentum of the multiplet member in the odd-$A$
nucleus.  Then, ECIS is used to find the deformation length $\delta_3$
that fits each multiplet member.  The deformation length for each
multiplet member is shown in Fig.~\ref{fig:e3_delta3}c.
Even the deformation lengths for the odd-A multiplet members are
remarkably constant near 1~fm, validating the weak-coupling
interpretation for these multiplets.

\section{V. Conclusion}

To conclude, the $(p,p')$ reaction in inverse kinematics has been used
to measure $^{48-52}$Ca. The energies and strengths of the octupole states
in these isotopes are remarkably constant with changing neutron
number, providing strong evidence that these octupole states are
dominated by proton excitations.  This is even true for the
odd-neutron $^{49,51}$Ca isotopes, in which the weak coupling of the odd
neutron to the core even-even nuclei appears to describe the observed
behavior well.  This is in strong contrast with the $2^+_1$ states, for
which the energies vary strongly with neutron number.  This behavior
is consistent with the $2^+_1$ states being dominated by neutron
excitations.

\begin{acknowledgments}
This work was supported by the National Science Foundation under Grant
Nos.  PHY-1303480, PHY-1064819, and PHY-1102511. GRETINA was funded by
the US DOE - Office of Science. Operation of the array at NSCL is
supported by NSF under Cooperative Agreement PHY-1102511(NSCL) and DOE
under grant DE-AC02-05CH11231(LBNL). We also thank T.J. Carroll for
the use of the Ursinus College Parallel Computing Cluster, which is
supported by NSF grant no. PHY-1205895.
\end{acknowledgments}

\end{document}